\documentclass[showpacs, pra,preprintnumbers ,amsmath, amssymb, superscriptaddress, aps]{article}%{revtex4-2}%[11pt]{article}
\usepackage{float}
\usepackage{array}
\usepackage{cool}
\usepackage{bbold}
\usepackage{eulervm}
\usepackage{hyperref}
\hypersetup{colorlinks=true,linkcolor=blue,citecolor=blue}
\let\oldmb\mathbold
\protected\def\mathbold{\oldmb}
\usepackage{amsmath}
\usepackage{todonotes}
\usepackage{graphicx}
\usepackage{subfig}
\usepackage{color}
 \usepackage{multicol}

\begin{document}
\title{\bf Quantum Hall Goos–Hänchen effect in graphene}
\author{D. Jahani \footnote{\href{d.jahani@sharif.edu}{d.jahani@sharif.edu}}, O. Akhavan, A. Alidoust Ghatar}
\maketitle {\it \centerline{
\emph{Department of Physics, Sharif University of Technology, P.O. Box 11155-9161, Tehran, Iran}}

%%% -------------------------

%%%%%%%%%%%%%%%%%%%%%%%%%%%%%%%%%%%%%%%%%%%%%%%%%%%%%%%%%%%%%%%%%%%%%%%%%%%%%%%%%%%%%%%%%%%%%%%%%%%%%%%%%%%%%%%%%%%%%%%%%%
%%% ABSTRACT %%%

\begin{abstract}
\emph{Strong Goos–Hänchen (GH) effect at a prism-graphene interface in the quantum Hall effect (QHE) condition is reported. Based on the full quantum description of the temperature-dependent surface conductivity of graphene present in the unconventional quantum Hall regime, magnetically strong tunable QHE GH shifts emerge. Our approach is based on deriving the generalized Fresnel coefficients with antisymmetric conductivity tensor for the Kerr phase of the incident linearly polarized light. Moreover, it is demonstrated that at low temperatures, GH shifts map plateaus as the intensity of the magnetic field grows. This quantum modulation of the GH effect in graphene by an applied magnetostatic bias may open doors to new opportunities for optical devices and QHE sensing applications in 2D materials.  }
\end{abstract}
\vspace{0.5cm} {\it \emph{Keywords}}: \emph{Graphene; Goos-Hänchen effect; Quantum Hall effect; Terahertz regime.}

\section{ Introduction}
\emph{A lateral shift occurs between an incident light beam and its predicted geometrical optical path for the reflected light at an interface when it undergoes the total internal reflection condition. This so-called Goos–Hänchen (GH) effect has been analyzed both theoretically and experimentally \cite{1,2,3}. This phenomenon has attracted immense attention due to the development of micro/nano-optics evolution at subwavelength scales \cite{4}. However, being just of the order of the wavelength of the incident light wave hinders its application area. Thus, the enhancement and control of GH effect have recently received rapidly increasing attention in various fields such as biomedicine, sensors, modulators and acoustics, quantum mechanics, and plasma physics \cite{5,6,7,8,9,10,11}. Up to now many ways such as layered structures, absorbing media, photonic crystals with and without a defect layer, metamaterials and prism coupling to 2D materials have been used to increase and control this effect \cite{12,13,14,15,16,17}. Moreover, GH effect in graphene, a single two-dimensional plane of carbon atoms in a honeycomb structure, has attracted great attention and, therefore, has already been the subject of some researches due to its extraordinary optical properties \cite{18,19,20}. For instance, many graphene-based structures including the Tamm states, photonic crystal (PC), and Prism-coupling have been introduced to raise GH effect in the terahertz region which is of most interest in sensing applications \cite{21,22,23,24,25,26,27}. }
\par
\emph{Now, the full description of the optical conductivity of graphene in QHE regime with optical excitation within the Landau levels (LLs) could be determined by the Kubo conductivity model. This QHE for optical transitions in single-layer graphene in the presence of a magnetostatic bias contributes as longitudinal and Hall components in its conductivity tensor which yields a Kerr effect rotation providing GH shifts of different polarization to the incident linearly polarized light upon the reflection. Thus, this temperature-dependent magneto-optical conductivity may produce GH effect which is large enough to be suitable for practical THz applications. }
\par
\emph{Also, the magneto-optical GH effect has been investigated in a prism-graphene coupling system with a magneto-optical Kerr effect(MOKE) for the incident polarized beam in Ref. \cite{28}. The authors in this paper, based on the classic model for the optical conductivity study the magneto-tunable enhanced GH shifts of graphene in the polar Kerr effect (PMOKE) configuration for which the magnetic field is perpendicularly applied on its surface. However, the enhancement of GH effect in graphene in their report is limited to the Drude model considering just the intra-band transitions. Now, in general, to study QHE GH effect in a carbonic single-layer structure like graphene it is essential to go beyond the semiclassical model and consider the full quantum description of its surface conductivity. In this work, we prove the existence of giant QHE GH shifts in a simple prism-graphene coupling device in QHE regime. Light-matter interaction in graphene under the unconventional QHE situation could bring giant GH shifts for different polarized light upon the total reflection. Hence, these giant QHE phase jumps in the reflected light for the introduced coupling structure could lead to dramatic improvement of THz GH effect in graphene. }
\par
\emph{The outline of the paper is divided into the following sections: The theoretical model and derivation of generalized Fresnel coefficients in QHE regime are introduced in Section \ref{2}. We consider graphene in the quantum QHE regime with optical transitions from LLs in Kubo conductivity model. In Section \ref{3} the numerical results and discussions are presented. The conclusion remarks are also addressed in Section \ref{4}. }
%%%%%%%%%%%%%%%%%%%%%%%%%%%%%%%%%%%%%%%%%%%%%%%%%%%%%%%%%%%%%%%%%%%%%%%%%%%%%%%%%%5
\section{Model and formula}\label{2}
\emph{In this work, we present the full quantum description of GH effect in a prism-graphene coupling system in QHE condition under the incident of $s$- and $p$-polarized light with the frequency $\omega$ as shown schematically in Fig. \ref{Fig. 1}. The dispersion of a linearly polarized beam passing through two dielectric media separated by monolayer graphene could be obtained by developing the components of the electric and magnetic vector of the incident, transmitted, and reflected beam as \cite{29}:
\begin{eqnarray}
\left\{\begin{array} {cc}
E_{i}=(-a_{p}\cos\theta_{i}, a_{s}, a_{p}\sin\theta_{i})e^{i\tau_{i}}\\
H_{i}Z_{0}=(-a_{s}n_{1}\cos\theta_{i}, -a_{p}n_{1}, a_{s}n_{1}\sin\theta_{i})e^{i\tau_{i}} \end {array} \right.
\end{eqnarray}
\begin{eqnarray}
\left\{\begin{array} {cc}
E_{r}=(-r_{p}\cos\theta_{r},r_{s}, r_{p}\sin\theta_{r})e^{i\tau_{r}}\\
H_{r}Z_{0}=(-r_{s}n_{1}\cos\theta_{r}, -r_{p}n_{1}e^{i\tau_{r}}, r_{s}n_{1}\sin\theta_{r})e^{i\tau_{r}} \end {array} \right.
  \end{eqnarray}
\begin{eqnarray}
\left\{\begin{array} {cc}
E_{t}=(-t_{p}\cos\theta_{t}, t_{s}e^{i\tau_{t}}, t_{p}\sin\theta_{t})e^{i\tau_{t}}\\
H_{t}Z_{0}=(-t_{s}n_{2}\cos\theta_{t}, -t_{p}n_{2}e^{i\tau_{t}}, t_{s}n_{2}\sin\theta_{t})e^{i\tau_{t}} \end {array} \right.
  \end{eqnarray}
with $a$, $r$, and $t$ indicating the complex amplitudes of the incidence, reflected, and transmitted light and $\tau_{m}=\omega t-\textbf{k}_{m}.\textbf{\textbf{r}}$ ($m=i,r,t$). Then, the electric and magnetic field components of the incident light at the interface, $z=0$, could be related by following equations:
\begin{eqnarray}
\left\{\begin{array} {cc}
E_{x}^{t}=E_{x}^{i}+E_{x}^{r} \ \ \ \ \ \ \ \ \ \ E_{y}^{t}=E_{y}^{i}+E_{y}^{r}, \\
H_{x}^{t}=H_{x}^{i}+H_{x}^{r}+J_{y}, \ \ \ \ \ \ \ \ \ \ H_{y}^{t}=H_{y}^{i}+H_{y}^{r}-J_{x} \end {array} \right.
  \end{eqnarray}
for which $J=\bar\sigma .E^{t}$ and $\bar\sigma$ are the surface current density and the optical conductivity of graphene, respectively. To proceed, in QHE condition, the surface conductivity of graphene develops a tensor which could be written in the following matrix form:
\begin{eqnarray}
\sigma=
\begin{pmatrix}
 \sigma_{L} & \sigma_{H} \\
 -\sigma_{H} & \sigma_{L}
\end{pmatrix}
\end{eqnarray}
As we use the quantum model for describing the optical transitions of LLs in a prism-graphene coupling system, the frequency-dependent longitudinal ($\sigma_{L}$) and Hall parts ($\sigma_{H}$) of its optical conductivity tensor are \cite{30}:
\begin{equation}\begin{split}
\sigma_{0}(\omega)=
&\frac{e^{2}v_{F}^{2}\vert eB\vert\left(\hbar\omega+2i\Gamma \right)}{\pi i}\times\\& \sum_{n=0}^{\infty}\left\lbrace\frac{\left[f_{d}(M_{n})-f_{d}(M_{n+1})\right]+ \left[f_{d}(-M_{n+1})-f_{d}(-M_{n})\right]}{\left(M_{n+1}-M_{n}\right)^{3}-\left(\hbar\omega+2i\Gamma \right)^{2}\left(M_{n+1}-M_{n}\right)}\right\rbrace \\&
+ \left\lbrace\frac{\left[f_{d}(-M_{n})-f_{d}(M_{n+1})\right]+ \left[f_{d}(-M_{n+1})-f_{d}(M_{n})\right]}{\left(M_{n+1}+M_{n}\right)^{3}-\left(\hbar\omega+2i\Gamma \right)^{2}\left(M_{n+1}+M_{n}\right)}\right\rbrace
 \end{split}\end{equation}
 and:
 \begin{equation}\begin{split}
&\sigma_{H}(\omega)=\frac{-e^{2}v_{F}^{2} eB}{\pi}\\&
\sum_{n=0}^{\infty}\left\lbrace \left[f_{d}(M_{n})-f_{d}(M_{n+1})\right]- \left[f_{d}(-M_{n+1})-f_{d}(-M_{n})\right]\right\rbrace \times \\&
\left\lbrace \frac{1}{\left(M_{n+1}-M_{n}\right)^{2}-\left(\hbar\omega+2i\Gamma \right)^{2}}+ \frac{1}{\left(M_{n+1}+M_{n}\right)^{2}-\left(\hbar\omega+2i\Gamma \right)^{2}}\right\rbrace
\end{split} \end{equation}
  where $E_{n}=\pm M_{n}=\sqrt{2n\vert eB\vert\hbar v_{F}^{2}}$ determines LLs and the scattering rate $\Gamma $ is assumed to be independent of the frequency of the light. The distribution function is $f_{d}(M_{n})=1/(1+exp[(M_{n}-\mu)/K_{B}T])$ with $e$, $\hbar$, $K_{B}$ and $ T$ representing the electron charge, reduced Planck constant, Boltzmann constant and the temperature, respectively.  In general, the existence of the nonzero Hall term in the surface conductivity of graphene indicates that the initial polarization of the incident light will not be preserved anymore for the reflected and transmitted light. Thus, a p-polarized (s-polarized) light could brings transmitted or reflected s-and p-polarized beam which could be shown by subscripts $sp$ ($ss$) and $pp$ ($ps$). }
\par
\emph{Now, we are able to provide QHE description of the emergent GH effect from reflected $ss$-, $sp$-, $ps$- and $pp$-polarized light based on the Fresnel equations for an incident linearly p- or s-polarized beam with the scattering angle, $\theta_{1}$ and the photon energy $E_{ph}$ in the THz region.  From equations (1)-(5) one can show that the reflection coefficients of $ss$-, $sp$-, $ps$- and $pp$-polarized beam are:
\begin{eqnarray}\label{(8)}
\left\{\begin{array} {cc} r_{ss}=-(1-\frac{2n_{1}f_{1}\cos\theta_{1}}{f_{1}f_{2}+z_{0}^{2}\sigma_{H}^2\cos\theta_{1}\cos\theta_{2}})a_{s},\\
r_{sp}=-(\frac{2n_{1}z_{0}\sigma_{H} \cos\theta_{1}\cos\theta_{2}}{f_{1}f_{2}+z_{0}^{2}\sigma_{H} ^2\cos\theta_{1}\cos\theta_{2}})a_{p},\\
r_{pp}=(1-\frac{2n_{1}f_{2}\cos\theta_{2}}{f_{1}f_{2}+z_{0}^{2}\sigma_{H} ^2\cos\theta_{1}\cos\theta_{2}})a_{p},\\
r_{ps}=-(\frac{2n_{1}z_{0}\sigma_{H} \cos\theta_{1}\cos\theta_{2}}{f_{1}f_{2}+z_{0}^{2}\sigma_{H} ^2\cos\theta_{1}\cos\theta_{2}})a_{s} \end {array} \right.
\end{eqnarray}
where $f_{1}$ and $f_{2}$ are defined as $f_{1}=n_{1}\cos\theta_{2}+n_{2}\cos\theta_{1}+z_{0}\sigma_{L}\cos\theta_{1}\cos\theta_{2}$ \ and\ $f_{2}=n_{1}\cos\theta_{1}+n_{2}\cos\theta_{2}+z_{0}\sigma_{L}$. Here, $z_{0}\approx 377 \Omega $ \ is the vacuum impedance. Note that, $a_{s}$\ and $a_{p}$ are the amplitude of the electric vector of the incident field in the perpendicular and parallel planes, respectively and the magnetic field is applied along the $z$ direction (see Fig. \ref{Fig. 1}). It is clear that for $a_{s}=a_{p}=1$ one gets $r_{sp}=r_{ps}$ and in the case of normal incidence $G_{1}=G_{2}$. Then, according to the stationary phase method, the amount of GH shifts could be obtained from:
\begin{eqnarray}
\Delta_{r_{ij}}=-\frac{\lambda}{2\pi}\frac{d\phi_{r_{ij}}}{d\theta}
\end{eqnarray}
where $\lambda=\frac{2\pi c}{\omega}$ is the wavelength of the incident beam and $\phi_{r_{ij}}$ ($r_{ij}$), with $i$ and $j$ indicating p and s, is the phases of the reflection coefficient ( reflection coefficient). At this point, it should be emphasized that our model is based on putting a layer of graphene on the prism in order to measure the GH shifts for the internal reflection condition in QHE regime. It is due the ease of the detection of the beams bringing the GH shift with high intensity and assuring that GH effect is not affected by any dip in the reflection before total reflection. Interestingly for lowest LL we have both electron and hole and therefore half-integer .
 Equation (9) could be written in the following form:
\begin{eqnarray}
\Delta_{r_{ij}}=-\frac{\lambda}{2\pi}\frac{1}{|r_{ij}|^{2}}\left(Re(r_{ij})\frac{d Im(r_{ij})}{d\theta}-Im(r_{ij})\frac{d Re(r_{ij})}{d\theta}\right)
\end{eqnarray}
 Note that, it is obvious that one can have $R=|r_{ij}|^{2}$ equal to unity when a beam of light hits an interface with indices of refraction $n_{1} > n_{2}$ under an angle $\theta_{i}= \theta_{c} =\arcsin  n_{2}/n_{1}$ at which geometrical optics predicts the total reflection. In the calculation of QHE GH shift, we treat the different components of the rotated polarized reflected light separately and calculate their contributions to each of QHE GH shifts. }
%%%%%%%%%%%%%%%%%%%%%%%%%%%%%%%%%%%%%%%%%%%%%%%%%%%%%%%%%%%%%%%%%%%%%%%%%%%%%%%%%%%%%%%%%%%%%%%%%%%%%%%%%%%%%%%%%%%
\section{Results and discussion}\label{3}
\emph{To satisfy the total internal reflection condition, we consider a prism with the refractive index $n=1.5$ coupled to a graphene layer with the surface optical conductivity tensor including nonzero Hall and longitudinal terms. In fact, from equation \ref{(8)} we see that what brings $sp$-polarized ($ps$-polarized) light in the reflection spectra of the introduced structure is the existence of a nonzero Hall term in the conductivity  which is dependent on the energy of the incident light, magnetic bias, chemical potential and the temperature. In our calculations the temperature is considered to be set as very low temperature $T=1\ K$ and also the room temperature $T=300\ K$ which is of most interest in the real-world THz application of optical devices. Also, note that in the classical model for GH effect in graphene we are not able to study the shifts at very low temperatures for which the unconventional QHE is observed \cite{31}.  }
\par
\emph{Fig. \ref{Fig. 2} (a-c) shows the dependence of the reflected beams versus the magnetic field and the chemical potential with different polarizations for the prism-graphene coupling device upon a specific incident angle for both incoming $s$- and $p$-polarized wave in the QHE situation. As it is clear, the total internal reflection for $R_{ss}$ and $R_{pp}$ occurs at the scattering angle $\theta=41.8104^{o}$. Interestingly, in this case, the reflection spectrum for  $R_{ps}= R_{sp}$ could be neglected. Furthermore, as is seen in Fig. \ref{Fig. 2} (d), at $\theta=41.8104^{o}$, one could observe giant QHE GH shift of $ss$-polarized wave for relatively high magnetic fields and the low temperature ($T=1\ K$) which is larger than the GH peak of $sp$/$ps$-polarized reflected light. Interestingly, Fig. \ref{Fig. 2} (e) shows a forbidden region for the presence of QHE GH shifts in lower area for the magnetic field intensity and high values of the chemical potential at $T=1\ K$. THe QHE GH spectrum for $pp$-polarized beam is shown in Fig. \ref{Fig. 2} (e) which reveals different results in comparison to the other polarized beam. }
\par
\emph{At very low temperature, $T=1\ K$, and for different photon energies Fig. \ref{Fig. 3} (a) reveals QHE GH shifts for $B=3.3\ T$ and $\mu=0.4\ eV$. It is observed that by growing the photon energy QHE GH shifts are decreased for $ss$-polarized light. This weak GH effect observed for stronger photon energies might be explained as that for stronger energies the surface conductivity of graphene shows smaller modulations in quantum regime. Note that the maximum value of the GH shift for this spectrum occurs at $E_{ph}=1\ meV$ and $\theta=41.8104^{o}$. Fig. \ref{Fig. 3} (b) illustrates both positive and negative QHE GH for $sp$ ($ps$) polarization of the reflected beam. We see that the largest positive GH shift similar to $ss$-polarized reflected beam appears at the same scattering angle $\theta=41.8104^{o}$. In addition, both positive and negative GH shifts could be neglected when the photon energy increases. However, in Fig. \ref{Fig. 3} (c), it is clear that the maximum amount of GH effect for $pp$-polarized beam appears at $\theta=41.8104^{o}$. Interestingly, the maximum value in this case shows to be almost independent of the photon energy. It is because at very low temperatures emergent QHE GH shifts of different frequencies show to be more localized in the frequency space.}
\par
\emph{In Fig. \ref{Fig. 4} we take our attention toward the effect of increasing the magnetic field on the GH effect for $\mu=0.4\ eV$ and $E_{ph}=1.5\ meV$. It is observed from Fig. \ref{Fig. 4} (a) that the peak of the $GH$ shift spectrum for $ss$-polarized light in the presence of an applied magnetic field appears at $\theta=41.8102^{o}$. However, it is seen that the GH peak of $sp$/$ps$-polarized light is much larger than the maximum of $ss$ polarized beam occuring at $\theta=41.8104^{0}$. It is clear that the spectrum in this case similar to the results of Fig. \ref{Fig. 4} (b), shows both positive and negative QHE GH shifts. In Fig. \ref{Fig. 4} (c) the maximum amount of QHE shifts for $pp$-polarized light waves are independent of the field intensity. }
\par
\emph{Note that, we neglect the effect of surface plasmon resonances (SPRs) which could occur in the total reflection condition for the introduced coupling structure. The reason is that SPRs decrease the amount of the reflected light in this condition as an observed dip in the reflection spectrum. Therefore, this dip in the reflection may result in observing larger GH shifts as it is clear from equation (10). However, the presence of an SPR indicates a dip in the reflection spectrum of the device weakening the reflected beam which could cause problem in the experimental detection of GH shifts. Therefore, there will be a trad-off between the enhancement of the GH shifts by exciting SPRs and the intensity of the reflected light. Therefore, all the shift peaks are induced by the total reflection of the incident linearly polarized beam at a specific scattering angle.}
\par
\emph{Fig. \ref{Fig. 5} shows the dependence of the QHE GH effect observed from $ss$-, $sp/ps$- and $pp$-polarized reflected beams on the intensity of the magnetic field for the incident angle $\theta=48.8104^{o}$. The variation of the GH effect spectrum is depicted in the field interval ranging from $1$ to $10\ (T)$ for different photon energy $E_{ph}$ (Fig. \ref{Fig. 5} (a-c)), chemical potential ($\mu$ Fig. \ref{Fig. 5} (d-f)) and temperature $T$ (Fig. \ref{Fig. 5} (g-i)). It is observed that for $\mu=0.4\ eV$, $T=1\ K$ and $E_{ph}=1.5\ meV$ the GH effect emerges as steps for stronger magnetic fields. This step-like feature of GH effect is due to the observation of  QHE for the conductance of graphene at applied strong magnetic fields and low temperatures. Also, it is seen that the modulation of GH shifts begin at a certain value for applied magnetic field which is observed in the GH effect which is due to Pauli blocking. Interestingly, as it is seen, at higher magnetic fields GH effect appears to be independent of the frequency at specific scattering angle. Moreover, by increasing the value of the chemical potential, as shown in Fig. \ref{Fig. 5} (d-f), QHE in prism-graphene structure reveals stronger GH effect for different polarized beams. Here, we see that strong QHE GH shift at low temperature, $T=1\ K$, can be observed for higher potential in the case of applying stronger magnetic fields. However, for lower field values as it is seen from the figure one observes  that increasing the chemical potential reveals weaker GH effect in graphene. We emphasize that this behaviour is not observed in the classical model considered to study GH effect in graphene. Additionally, as the temperature is considered to be $T=300\ K$, the step-like behavior tends to be disappeared. This expected behaviour of GH effect at higher temperatures which is shown in Fig. \ref{Fig. 5} (g-i), reflects the distribution of Dirac electrons of graphene in LLs levels. In fact, incrasing the temperature disturb the well separated LLs of graphene. Superisingly, the amount of GH effect at the room temperature remains almost the same relative to its low temperature values for $ss$, $sp/ps$ and decreases for $pp$-polarized reflected light. This might be of most interest in real-world sensing applications of GH effect-based optical devices which are considered to operate at the room temperature.  }
\par
 \emph{Now, we examine the QHE lateral shifts for $ss$-, $sp/ps$- and $pp$-polarized beams in the photon energy interval ranging from $E_{ph}=1\ meV$ to $E_{ph}=10\ meV$ as shown in Fig. \ref{Fig. 6} for the fixed scattering angle $\theta=48.8104^{o}$. In Fig. \ref{Fig. 6} (a-c), we have shown that by increasing the photon energy, the value of QHE GH shift is decreased gradually. In Fig. \ref{Fig. 6} (d-f) the behaviour of QHE GH shifts are depicted for different chemical potentials which show that one could gain the giant GH shift (1500) in lower frequencies. In Fig. \ref{Fig. 6} (g-i), GH shifts are presented for two different temperatures and fixed chemical potential. It is seen that the results at the low and room temperature show the same GH effect for different polarized reflected lights. }
\par
 \emph{As we see relative to the results reported in the case of considering GH effect for classical model of prism-graphene coupling which gives at most shifts equal to $100$,  one could obtain giant GH shifts near $1500$ for the structure under QHE situation. Moreover, it is revealed that incident $s$-polarized light shows much larger GH effect in quantum Hall model for the introduced structure. The story behind this giant QHE GH effect is that the optical conductivity of graphene under the influence of a normal magnetic bias marks much larger values than its classical counterpart. Moreover, the detection of individual quanta by GH effect in graphene requires quantum methods to be employed in studying the sensing properties of this 2D carbonic material. Such sensitivity is beyond the reach of any detection method which employs a classical model. Therefore, tunable GH shifts in QHE regime could be used for detection of low magnetic field intensity and also for THz applications in graphene-based sensing devices.}
 \par
\emph{time.}

%%%%%%%%%%%%%%%%%%%%%%%%%%%%%%%%%%%%%%%%%%%%%%%%%%%%%%%%%%%%%%%%%%%%%%%%%%%%%%%%%%
\section{Conclusion}\label{4}
\emph{In this work, we prove the existence of an adjustable model for achieving giant GH shifts from different polarized reflect light in a prism-graphene coupling device under the QHE regime. We show that giant tunable GH shifts in graphene could be induced upon the total reflection of the incident polarized light waves at a specific scattering angle. The QHE GH shifts are studied by deriving generalized Fresnel coefficients for the incident linearly polarized light including an antisymmetric conductivity tensor realized by applying a constant magnetic field on graphene's surface. Significantly, GH shifts undergo step-like changes at the low temperature and strong magnetic fields which diminish at higher temperatures. Here, we notice that QHE in graphene could be also observed at the room temperature \cite{32}. As graphene is an one-atom-thick 2D carbonic material quantum Hall GH effect could suggest very sensitive tool in THz sensing applications. This is far beyond the reach of any detection methods for which the classical models for studying GH effect in graphene are introduced. In the end, we note that introduced Fresnel reflection coefficients of the graphene-prism coupling structure in the presence of an applied magnetostatic field is different from those suggested in \cite{33,34}. Therefore, our numerical results for QHE GH effect are different from results have been reported for GH shifts.  }
\section{Appendix: Fresnel reflection coefficients}
\emph{Here, we present a detailed calculation of the reflected spectrum for the introduced structure and show that it could recover equation from the following expression introduced in \cite{33,34}.
\begin{eqnarray}
r_{ss}=-\frac{\alpha_{-}^T\alpha_{+}^L+\beta}{\alpha_{+}^T\alpha_{+}^L+\beta},\ \ r_{pp}=\frac{\alpha_{+}^T\alpha_{-}^L+\beta}{\alpha_{+}^T\alpha_{+}^L+\beta},\ \ r_{sp}=r_{ps}=-2\sqrt{\frac{\mu_{0}}{\epsilon_{0}}}\frac{k_{1z}k_{2z}\sigma_{H}}{\alpha_{+}^T\alpha_{+}^L+\beta}
\end{eqnarray}
where
\begin{eqnarray}
\alpha_{\pm}^L=(k_{1z}\epsilon_{2}\pm k_{2z}\epsilon_{1}+k_{1z}k_{2z}\frac{\sigma_{L}}{\epsilon_{0}\omega}),\ \ \alpha_{\pm}^T=(k_{2z}\pm k_{1z}+\mu_{0}\omega\sigma_{T}),\ \ \beta=\frac{\mu_{0}}{\epsilon_{0}}k_{1z}k_{2z}\sigma_{H}^2
\end{eqnarray}
Now by the use of relation $K_{iz}=\frac{\omega}{c}n_{i}\cos(\theta_{i})$ we can write:
\begin{eqnarray}
\alpha_{\pm}^L=\frac{\omega}{c}n_{1}n_{2}[n_{2}\cos\theta_{1}\pm n_{1}\cos\theta_{2}+\frac{1}{c\epsilon_{0}}\sigma_{L}\cos\theta_{1}\cos\theta_{2}],\ \ \alpha_{\pm}^T=\frac{\omega}{c}[n_{2}\cos\theta_{2}\pm n_{1}\cos\theta_{1}+\mu_{0}c\sigma_{T}],\ \
\end{eqnarray}
and
\begin{eqnarray}
\beta=\frac{\omega^2}{c^2}n_{1}n_{2}\frac{\mu_{0}}{\epsilon_{0}}\sigma_{H}^2\cos\theta_{1}\cos\theta_{2}
\end{eqnarray}
At this point one might recognize that:
\begin{eqnarray}
\alpha_{-}^L=\alpha_{+}^L-2\frac{\omega}{c}n_{1}n_{2}n_{1}\cos\theta_{2},\ \ \ \ \alpha_{-}^T=\alpha_{+}^T-2\frac{\omega}{c}n_{1}\cos\theta_{1}
\end{eqnarray}
Then with the help of following relations for the impendance of free space $z_{0}$:
\begin{eqnarray}
Z_{0}=\sqrt{\frac{\mu_{0}}{\epsilon_{0}}}=\mu_{0}c=\frac{1}{c\epsilon_{0}}
\end{eqnarray}
one could proceed one step forward and write:
\begin{eqnarray}
\alpha_{+}^L=\frac{\omega}{c}n_{1}n_{2}f_{1}, \ \ \  \alpha_{+}^T=\frac{\omega}{c}f_{2},\ \ \beta=\frac{\omega^2}{c^2}n_{1}n_{2}z_{0}^2\sigma_{H}^2\cos\theta_{1}\cos\theta_{2}
\end{eqnarray}
with the definitions:
\begin{eqnarray}
f_{1}=[n_{2}\cos\theta_{1}+ n_{1}\cos\theta_{2}+z_{0}\sigma_{L}\cos\theta_{1}\cos\theta_{2}], \ \ f_{2}=[n_{2}\cos\theta_{2}+ n_{1}\cos\theta_{1}+z_{0}\sigma_{T}]
\end{eqnarray}
In the end, by substituting equations (15) and (17) in (11)  Fresnel reflection coefficients of the
graphene-prism coupling system in the presence of a constant magnetic field applied on its surface is given by:
\begin{eqnarray}\label{(8)}
\left\{\begin{array} {cc} r_{ss}=-(1-\frac{2n_{1}f_{1}\cos\theta_{1}}{f_{1}f_{2}+z_{0}^{2}\sigma_{H}^2\cos\theta_{1}\cos\theta_{2}}),\\
r_{sp}=r_{ps}=-(\frac{2n_{1}z_{0}\sigma_{H} \cos\theta_{1}\cos\theta_{2}}{f_{1}f_{2}+z_{0}^{2}\sigma_{H} ^2\cos\theta_{1}\cos\theta_{2}}),\\
r_{pp}=(1-\frac{2n_{1}f_{2}\cos\theta_{2}}{f_{1}f_{2}+z_{0}^{2}\sigma_{H} ^2\cos\theta_{1}\cos\theta_{2}}) \end {array} \right.
\end{eqnarray}
which are almost the same as we have derived independently from the relations (1)-(5). However, to be exactly recovered one instead requires to use the following relations:
\begin{eqnarray}
r_{ss}=-\frac{\alpha_{-}^L\alpha_{+}^L+\beta}{\alpha_{+}^L\alpha_{+}^L+\beta},\ \ r_{pp}=\frac{\alpha_{+}^L\alpha_{-}^L+\beta}{\alpha_{+}^L\alpha_{+}^L+\beta},\ \ r_{sp}=r_{ps}=-2\sqrt{\frac{\mu_{0}}{\epsilon_{0}}}\frac{k_{1z}k_{2z}\sigma_{H}}{\alpha_{+}^L\alpha_{+}^L+\beta}
\end{eqnarray}
with
\begin{eqnarray}
f_{1}=[n_{2}\cos\theta_{1}+ n_{1}\cos\theta_{2}+z_{0}\sigma_{L}\cos\theta_{1}\cos\theta_{2}], \ \ f_{2}=[n_{2}\cos\theta_{2}+ n_{1}\cos\theta_{1}+z_{0}\sigma_{L}]
\end{eqnarray}
}
%%%%%%%%%%%%%%%%%%%%%%%%%%%%%%%%%%%%%%%%%%%%%%%%%%%%%%%%%%%%%%%%%%%%%%%%%%%%%%%%%%%%%%%%%%%%%%%%%%%%%%%%%%%%%%
\section{Data availability}
Data sharing does not apply to this article as no new data were created or analyzed in this study.
\section{Acknowledgment}
A. Alidoust Ghatar is an invited researcher by D. Jahani
%%%%%%%%%%%%%%%%%%%%%%%%%%%%%%%%%%%%%%%%%%%%%%%%%%%%%%%%%%%%%%%%%%%%%%%%%%%%%%%%%%%%%%%%%%

\
\pagebreak
\

\begin{figure}[b!]
\begin{center}
\includegraphics[width=18cm]{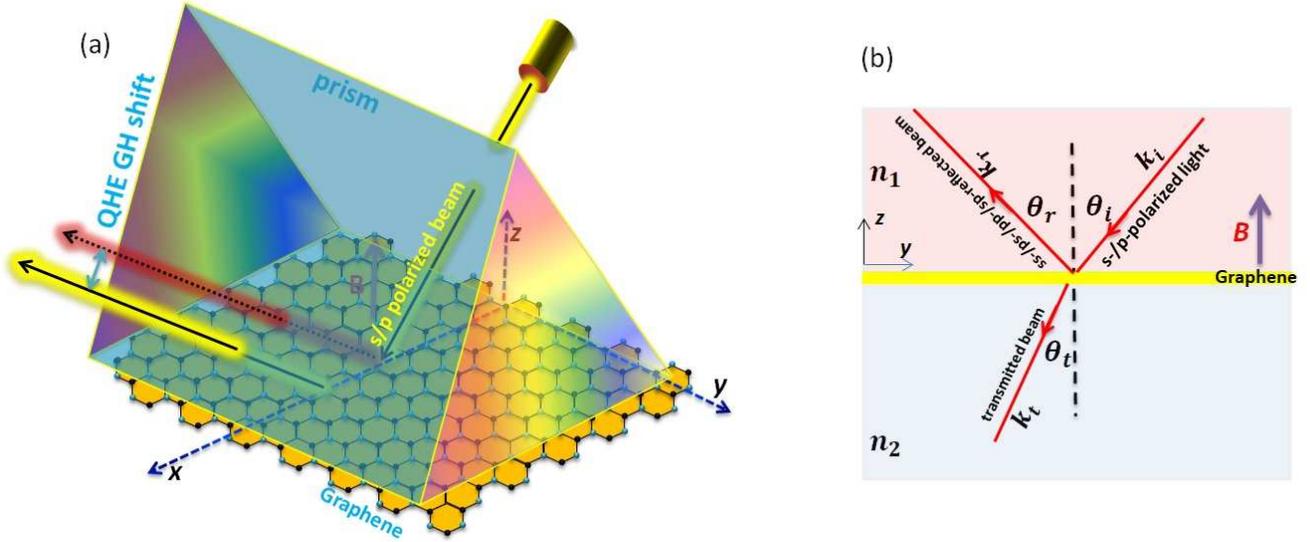}
 \caption{(a) Schematic representation of the GH shift of the designed structure in the presence of a prism with the refractive index of $n=1.5$ and a graphene layer In QHE regime. (b) Schematic view of the scattering of an incident lineally polarized light between two media with refractive indices $n_{1}$ and $n_{2}$ separated by a graphene layer under an applied magnetic field $B$. (A colour version of this figure can be viewed online.)} \label{Fig. 1}
\end{center}
\end{figure}

\begin{figure}[b!]
\begin{center}
\includegraphics[width=18cm]{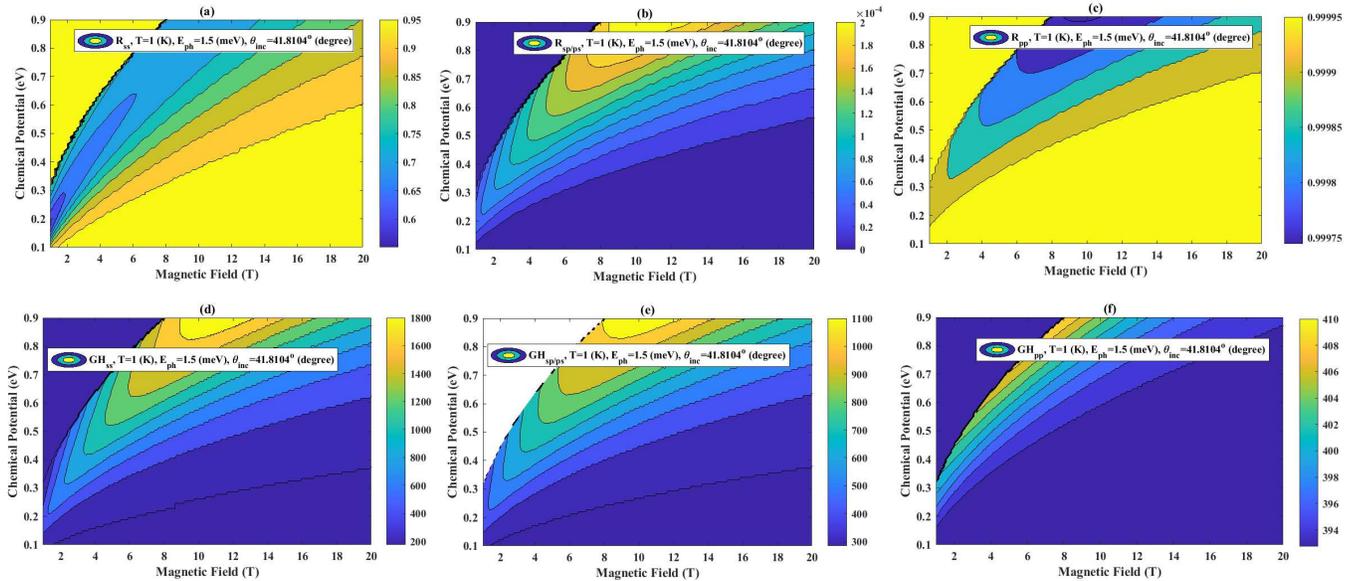}
 \caption{The QHE GH shifts of $ss$-, $ps$/$sp$- and $pp$-polarized light for the magnetic field interval $B=1-45\ T$ in the case of three different values of the chemical potential with the photon energy $E_{ph}=1.5\ meV$. (A colour version of this figure can be viewed online.)}\label{Fig. 2}
\end{center}
\end{figure}

\begin{figure}[b!]
\begin{center}
\includegraphics[width=18cm]{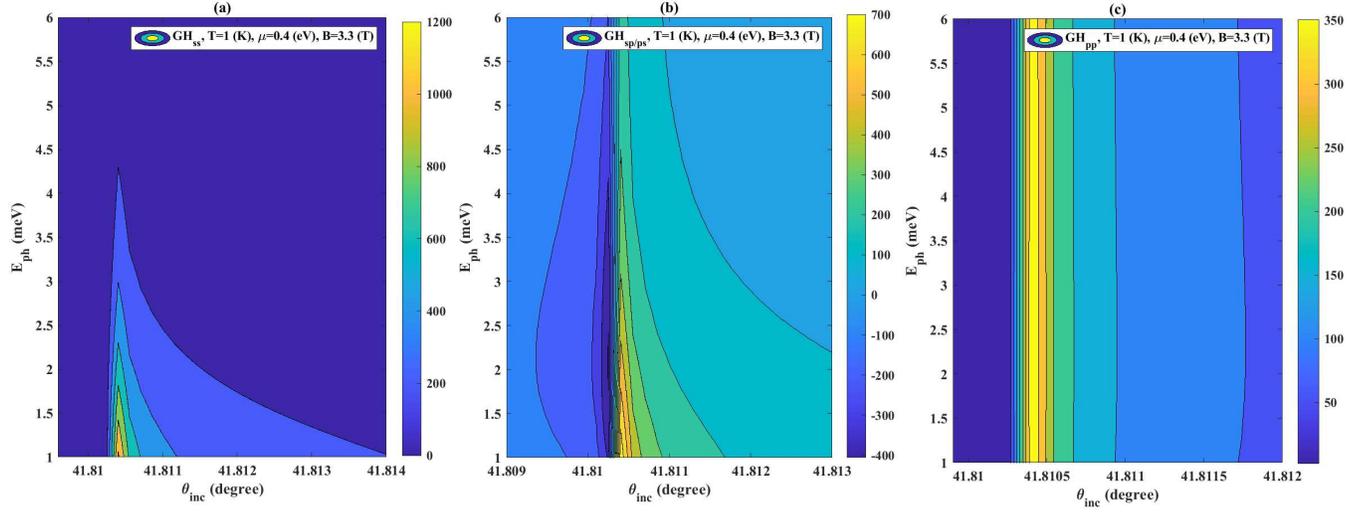}
 \caption{The QHE GH effect versus with different photon energies and scattering angles at $T=1\ K$ for $\mu=0.4$ and $B=3.3\ T$. (A colour version of this figure can be viewed online.) }\label{Fig. 3}
\end{center}
\end{figure}

\begin{figure}[b!]
\begin{center}
\includegraphics[width=18cm]{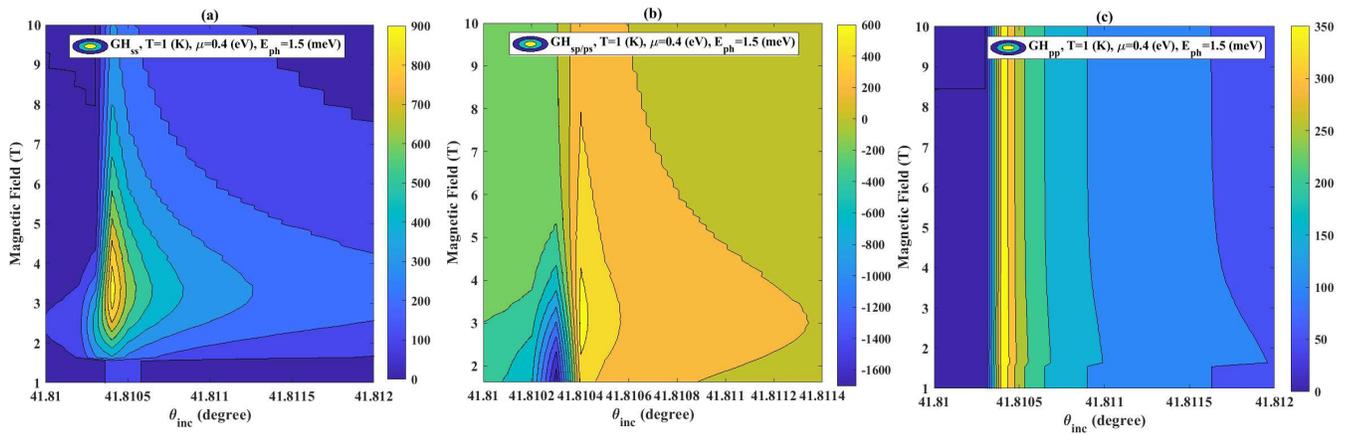}
 \caption{Magneto-optic modulation of GH effect in the prism-graphene coupling structure for different values of the scattering angle. (A colour version of this figure can be viewed online.)}\label{Fig. 4}
\end{center}
\end{figure}

\begin{figure}[b!]
\begin{center}
\includegraphics[width=19cm]{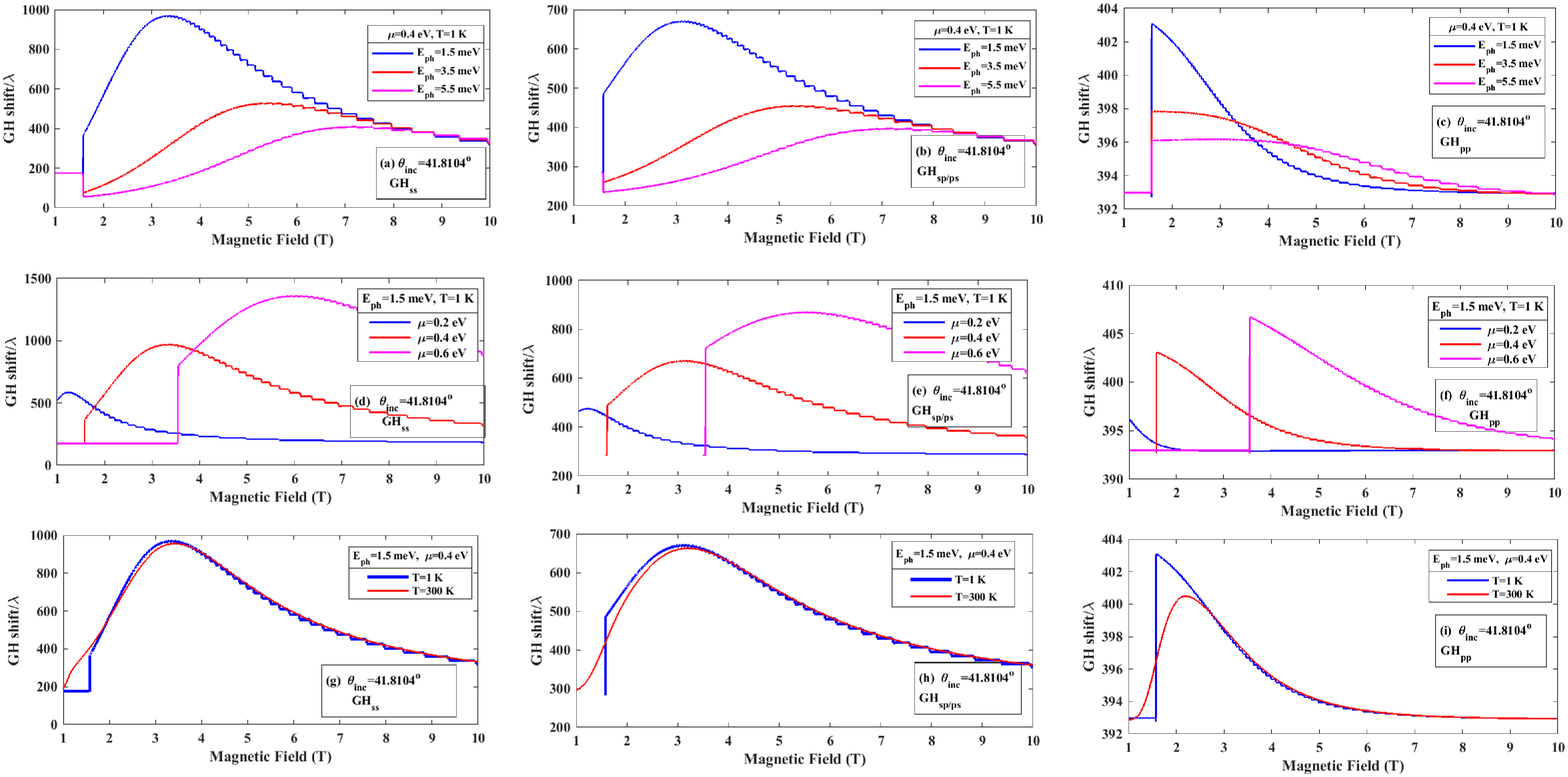}
 \caption{Modulation of QHE GH shifts versus the magnetic field for: (a-c) different photon energies (d-f) different chemical potentials; (g-i) different scattering angles. (A colour version of this figure can be viewed online.)}\label{Fig. 5}
\end{center}
\end{figure}

\begin{figure}[b!]
\begin{center}
\includegraphics[width=19cm]{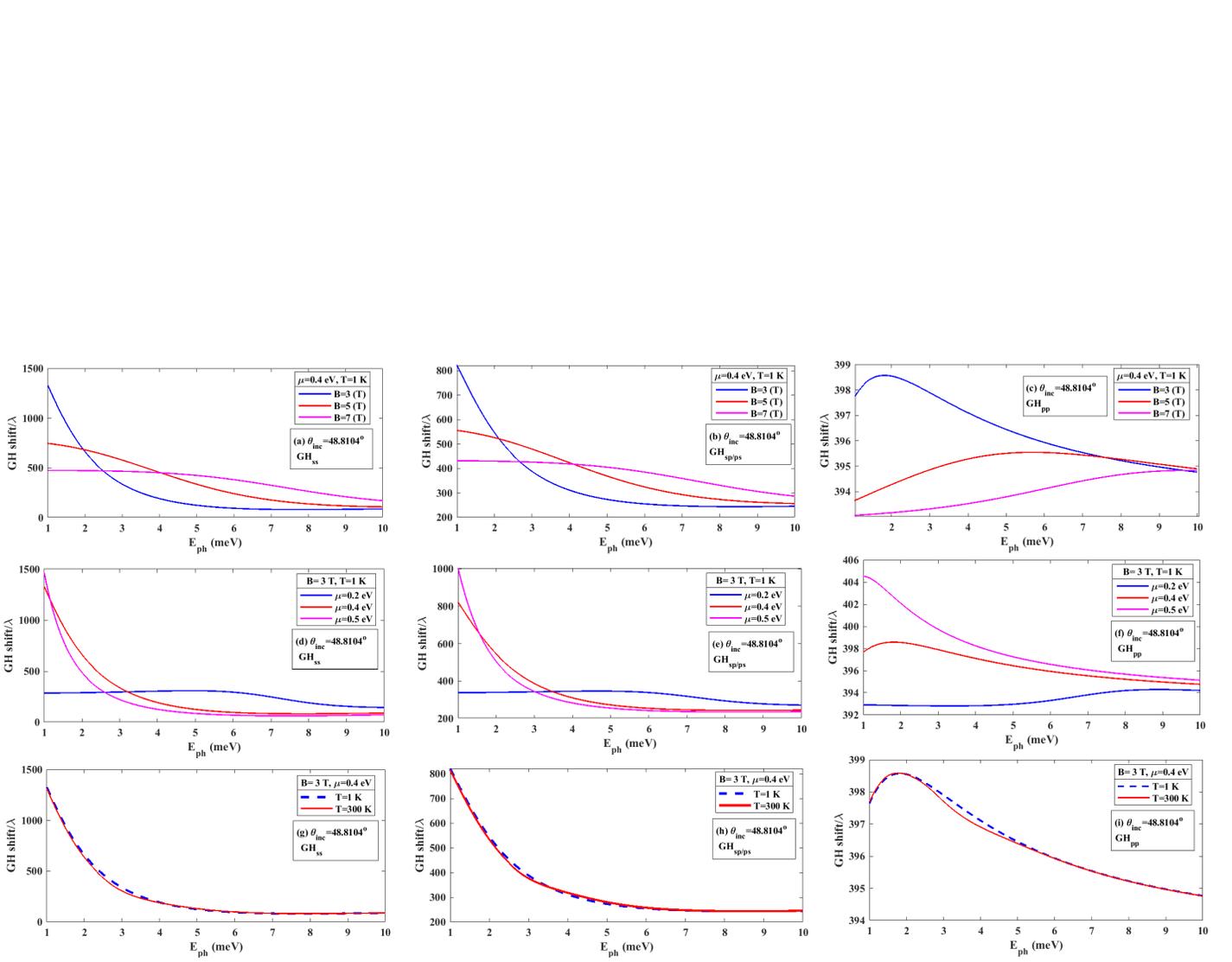}
 \caption{ Modulation of QHE GH shifts versus the photon energy for: (a-c) different magnetic fields (d-f) different chemical potentials; (g-i) two different temperatures. (A colour version of this figure can be viewed online.)}\label{Fig. 6}
\end{center}
\end{figure}

%%%%%%%%%%%%%%%%%%%%%%%%%%%%%%%%%%%%%%%%%%%%%%%%%%%%%%%%%%%%%%%%%%%%%%%%%%%%%%%%%%%%%%%%%%%%%%%%%%%%%%%%%%%%%%%%%%%%%%

\end{document}